\documentclass[
    pra,
    aps,
    twocolumn,
    superscriptaddress,
    notitlepage,
    10pt,
 final,   
]{revtex4-1}

\usepackage{graphicx}
\usepackage{xcolor}
\usepackage{ragged2e}
\usepackage{ifdraft}
\usepackage{lineno}
\usepackage{booktabs}
\usepackage{lipsum} 
\usepackage{titletoc}
\titlecontents{subsubsection}
              [3.3em]
              {\footnotesize\protect\addvspace{5pt}}%
              {\thecontentslabel~~}
              {}%
              {\hfill\contentspage}%

\usepackage{amsmath, amssymb, amsthm, bm}
\usepackage{physics, physconst, physunits}
\usepackage{braket}

\usepackage{pstricks}
\usepackage{tikz, tkz-orm, pgfplots}
\usetikzlibrary{arrows,shapes,backgrounds,decorations.markings}
\usetikzlibrary{matrix,positioning,decorations.pathreplacing,calc,tikzmark}
\pgfplotsset{width=7.5cm,compat=1.16}
\usepgfplotslibrary{fillbetween}

\usepackage{hyperref, bookmark}
\hypersetup{hidelinks} 
\usepackage{appendix}

\ifdraft{

    \linenumbers
}{}

\renewcommand{\iota}{i}

\usepackage{bbm}
\usepackage{amsmath}
\usepackage{mathrsfs}
\usepackage[normalem]{ulem}
\usepackage{float}
\usepackage{graphicx}
\usepackage{soul}
\def\lket#1{\vert#1\rangle\hspace{-1mm}\rangle}
\def\llket#1{#1\rangle\hspace{-1mm}\rangle}
\def\lbra#1{\langle\hspace{-1mm}\langle#1\vert}

\begin{document}
\title{Symmetries \& Correlations in Continous Time Crystals}
\author{Ankan Mukherjee}
\affiliation{Department of Physics, Indian Institute of Technology-Bombay, Powai, Mumbai 400076, India}
\author{Yeshma Ibrahim}	
\affiliation{Department of Physics, Indian Institute of Technology-Bombay, Powai, Mumbai 400076, India}
\email{yeshma@phy.iitb.ac.in}
\author{Michal Hajdu\v{s}ek}
\affiliation{Keio University Shonan Fujisawa Campus, 5322 Endo, Fujisawa, Kanagawa 252-0882, Japan}
\affiliation{Keio University Quantum Computing Center, 3-14-1 Hiyoshi, Kohoku, Yokohama, Kanagawa 223-8522, Japan}
\author{Sai Vinjanampathy}
\email{sai@phy.iitb.ac.in}
\affiliation{Department of Physics, Indian Institute of Technology-Bombay, Powai, Mumbai 400076, India}
\affiliation{Centre of Excellence in Quantum Information, Computation, Science and Technology, Indian Institute of Technology Bombay, Powai, Mumbai 400076, India.}
\affiliation{Centre for Quantum Technologies, National University of Singapore, 3 Science Drive 2, Singapore 117543, Singapore.}
\begin{abstract}
    We demonstrate the inadequacy of mean-field theory by exploring the effects of initial state correlations on the dynamics of continuous time crystals, necessitating higher-order cumulant expansions. We exemplify this using cat states for which the mean field fails to predict a phase transition but the second order cumulant expansion theory captures it. Motivated by the symmetries of the system, we choose a truncation of cumulant theory at the second-order and demonstrate that it is sufficient to accurately capture the dynamical features overlooked by the mean-field.
\end{abstract}

\maketitle
 
\section{Introduction}

Time crystals are understood to break either discrete time translation symmetry or continuous time translation symmetry. Of interest to us here, are continuous time crystals (CTCs)~\cite{sacha2017time, else2020discrete, khemani2019brief} that have been proposed in open quantum systems, where the time crystal regime is stabilized by the presence of dissipation. 
These models exhibit a dissipative phase transition as the ratio of drive strength to dissipation crosses unity, resulting in non-analytic changes in the steady states of open quantum systems~\cite{PhysRevA.Minganti}.
Examples of such CTCs are non-interacting collective models such as the driven Dicke model~\cite{mattes_entangled_2023, PhysRevLett.121.035301}, interacting collective models like the $p$ - $q$ model~\cite{piccitto_symmetries_2021}, power-law decaying spin models~\cite{PhysRevB.106.224308} and central spin models~\cite{ krishna_measurement-induced_2023}.
There has been extensive theoretical analysis on several aspects of these models including their symmetries, multipartite correlations and the underlying phase transition~\cite{piccitto_symmetries_2021, lourenco_genuine_2022}.
Recent experiments have observed CTCs in an atomic Bose-Einstein condensate~\cite{kongkhambut_observation_2022}, and demonstrated remarkably long-lived broken-symmetry phase in electron-nuclear spin system in a semiconductor~\cite{greilich2024robust}.

Typical method to study the dynamics of such open quantum system models, given by the Gorini–Kossakowski–Sudarshan–Lindblad (GKSL) master equation, is to analyze the generator in Liouville space~\cite{breur, manzano}. 
CTCs can be effectively studied within this Liouville superoperator formalism. 
Since a complete description of open system dynamics involves an increasingly intractable exact diagonalization of the Liouville superoperator, it becomes necessary to resort to a mean field approximation to qualitatively study the systems for large system sizes.  
However, unlike typical phase transitions, the contributions of multiple eigenvalues in the steady state can induce an initial state dependence on the dynamics of boundary time crystals producing qualitatively different behaviors. 
Recently, initial state dependence has been found to have a profound impact on the behavior of time crystals~\cite{solanki2024exotic}, all within the purview of mean-field approximation.

We investigate the effect of correlated initial states on the behavior of continuous time crystals. Of particular interest to us is the validity of mean-field approximations in the presence of initial correlations. 
While for initially uncorrelated states, mean-field description is exact~\cite{mori2013exactness, carollo2021exactness}, it might become insufficient for initially correlated states. 
We address this issue by resorting to cumulant expansion theories that can enable a comparatively computationally inexpensive description for the dynamics of initially correlated states~\cite{Kubo_original, PhysRevA.82.033810, PhysRevLett.102.163601, PhysRevA.78.022102, kira2011semiconductor, kirton2018superradiant, PhysRevA.100.013856}. 
This method casts the evolution of $n$-particle observables (the $n$th order) into a quantum analogue of Bogoliubov–Born–Green–Kirkwood–Yvon (BBGKY) hierarchy, where the evolution of each order depends on terms up to the immediate next order. 
We attempt to reasonably approximate the dynamics of continuous time crystals with correlated initial states, and thereby demonstrate the usefulness of cumulant expansion theory, especially in regimes where mean field theory fails. Indeed, initial studies on the stability of time crystals~\cite{watanabe_absence_2015} following the original suggestion by Wilczek~\cite{wilczek_quantum_2012} also involved correlation functions in space and time, which we take as additional motivation for this work.

The paper is organized as follows. In Section~\ref{sec: btc}, we introduce continuous time crystals and briefly discuss its key dynamical features within open quantum systems theory. Section~\ref{sec: limits} is used to illustrate the effects of correlated initial states on the dynamics of time crystalline systems within the mean-field approximation. Section~\ref{sec: cumulants} encapsulates cumulant expansion theory and discusses it in the context of continuous time crystals. The key results of our study are presented in Section~\ref{sec: results}. We further provide a discussion of our findings and conclude in Section~\ref{sec: conclusion}.

\section{Continuous Time Crystals}\label{sec: btc}
Continuous time crystals form a class of open quantum system models, the steady states of which spontaneously break the continuous time translational invariance in the thermodynamic limit~\cite{iemini_boundary_2018}.
They possess a time-independent drive along with dissipation that is characteristic to open quantum systems.
The emergence of continuous time crystals is a purely self-organized process that can be thought of as a dissipative phase transition arising from the many-body correlations due to a competition between coherent drive and dissipation.

Spin models, both interacting and non-interacting, have been archetypal systems for studies on these continuous time crystals~\cite{piccitto_symmetries_2021}. They are studied within an open quantum system formalism where the evolution of the density matrix is described by the GKSL master equation. 
In the canonical model of continuous time crystals, the dynamics is generated by the following Liouville superoperator $\mathscr{L}$ in the Lindblad form
\begin{equation}\label{eq:Lindy}
    \dot{\rho} = \mathscr{L}[\rho] = -i[\Omega \hat{S}_{x}, \rho] + \frac{\kappa}{S}\mathscr{D}[\hat{S}_{-}]\rho,
\end{equation}
where $\mathscr{D}[\hat{X}]\rho = \hat{X}\rho \hat{X}^{\dagger} - 1/2\{\hat{X}^{\dagger}\hat{X}, \rho\}$ is the dissipator with Lindblad operator $\hat{X}$. 
In the above equation, $S = N/2$ is the total spin of a system with $N$ atoms and $\hat{S}_{\alpha} = \sum_{i} \sigma_{\alpha}^{i}$, $\alpha \in \{x,y,z\}$, are the collective spin operators where $\hat{S}_{\pm} = (\hat{S}_{x} \pm i \hat{S}_{y})/2$. 
The unitary evolution here is given by the drive Hamiltonian $\hat{S}_{x}$ with strength $\Omega$, while the dissipation is modeled by $\hat{S}_{-}$ with strength $\kappa$. 
Such a system possesses the strong symmetry, $\hat{S}^2$, that results in the dynamics being confined to the maximally polarized subspace~\cite{piccitto_symmetries_2021}. 
In the regime where drive strength exceeds dissipation $(\Omega/\kappa > 1)$, the system exhibits oscillations that decay to a time-independent steady state on a time scale that diverges with increasing system size. 
As a consequence, in the thermodynamic limit, these oscillations become persistent and are witnessed by the order parameter $\langle \hat{S_z}\rangle/S$.
On the other hand, in the regime of strong dissipation $(\Omega/\kappa \leq 1)$, the system quickly decays to a stationary state with a constant value of $\langle \hat{S_z}\rangle/S$. 
These regimes and the emergence of a boundary time crystal can be better understood through a rigorous study of the properties of the Liouville superoperator given by Eq.~(\ref{eq:Lindy}), which we briefly review for completeness. 

Using the Fock-Liouville space of vectorised density matrices, any Markovian evolution in the Lindblad form can be expressed as $\lket{\dot{\rho}} = \hat{\mathcal{L}}\lket{\rho}$~\cite{manzano, AlbertPRA}. 
Here, $\hat{\mathcal{L}}$ is the Liouville superoperator in matrix form and $\lket{\rho}$ is the vectorised density matrix.
The Liouvillian can be expressed in the spectral form as $\hat{\mathcal{L}} = \sum_{k}\lambda_k \lket{r_k}\lbra{l_k}$, where $\lambda_k = \alpha_k + i\beta_k$ denote complex eigenvalues with corresponding left and right eigenvectors, $\lket{l_k}$ and $\lket{r_k}$. 
Since $\hat{\mathcal{L}}$ is non-Hermitian, the left and the right eigenvectors may be different. 
We note that the complex eigenvalues come in conjugate pairs with non-positive real parts, and $\hat{\mathcal{L}}$ has at least one zero eigenvalue \cite{krishna2023select}. 
These eigenvalues determine the dynamical behavior of the system.
Eigenvalues with a non-zero real part are transients and contribute little to the long time dynamics of the system.
The case of $\alpha_k \neq 0$ and $\beta_k = 0$ results in an exponential decay to a stationary state, while the outcome of $\alpha_k \neq 0$ and $\beta_k = 0$ is a spiraling decay. 
On the other hand, the eigenvalues with $\alpha_k = 0$ determines the steady state properties of a system. The $\alpha_k = 0$, $\beta_k = 0$ corresponds to the stationary states, and the purely imaginary eigenvalues with $\alpha_k = 0, \beta_k \neq 0$ result in persistent oscillations at frequency $\beta_k$.  

Within the formalism discussed above, the time crystal regime of the boundary time crystal model ($\Omega/\kappa > 1$) can be studied using a finite size scaling of the real and complex parts of the eigenvalues. 
Such scaling has revealed that as the system size $N$ goes to infinity, the complex eigenvalues close in towards the imaginary axis, resulting in a steady state with persistent oscillations in the thermodynamic limit  --- a picture that is consistent with the idea of dissipative phase transitions with closing spectral gap~\cite{PhysRevA.Minganti, hajdusek_seeding_2022, PhysRevA.86.012116}. 
This phase transition is characterized by non-analytic changes in the steady-state behavior of order parameters in the thermodynamic limit. 
However, the exact diagonalization of the Liouville superoperator becomes a formidable task in the thermodynamic limit. This demands approximate techniques like the mean-field description that improves the computational resources required and reduces the problem to solving a small set of differential equations.

\section{Limitations of Mean Field Analysis} \label{sec: limits}
As noted before, an exact simulation of quantum many-body systems becomes intractable with growing system size owing to the exponential increase in the Hilbert space dimensionality. 
The presence of symmetries often helps to reduce the effective dimensionality of the system and enables a complete description~\cite{gross1996role, manzano2014symmetry, article2022symmetry, buvca2012note, AlbertPRA}. 
However, any approach that attempts to describe a generic many-body system in terms of fewer variables can only be approximate. 
The spirit of such approximations usually entails neglecting higher order correlations in the system, and the exactness of this method has been a field of active study~\cite{PhysRevA.93.023821, PhysRevX.6.031011, mori2013exactness, carollo2021exactness, fiorelli2023mean}. 

An example of this is mean-field theory that seeks to depict a many-body system in terms of a small set of effective single-particle observables. 
Such a mean-field description has been proven to be exact for product initial states in the case of collective models like the open Dicke model~\cite{mori2013exactness, carollo2021exactness}. Continuous time crystals themselves have been successfully studied within this approximation for initial states such as the well known extremal Dicke state, $\ket{S,S}$, and the spin coherent state that are amenable to a successful mean-field description as the higher order correlations for these states vanish in the thermodynamic limit (see Appendix~\ref{app: cum_for_cohstate}). 
However, according to Levy's lemma, any generic pure state in a high-dimensional Hilbert space is expected to be strongly entangled~\cite{milman1986asymptotic, ledoux2001concentration}. 
Therefore, the study of collective models like the continuous time crystal initialized to generic pure initial states are limited both by the general computational complexity in dealing with high dimensional systems, combined with a failure of such mean-field descriptions. 

To illustrate this limitation, we present the case of a class of entangled states called the $k-$uniform states, where $k \leq N/2$~\cite{k-unikarol,PRR-kuni}. For an $n$-qubit state $\ket{\psi} \in \mathcal{H}^{\otimes n}$, where $\mathcal{H}:= \mathbbm{C}^2$ is the local Hilbert space corresponding to the individual qubits, a $k-$uniform state is defined as those states for which  
\begin{equation}
    \rho_S = \text{Tr}_{\Bar{S}}\left( \ket{\psi}\bra{\psi}\right) \propto \mathbbm{1}, \quad \forall S \subset \left\{ 1, ..., n\right\}, |S|\leq k,
\end{equation}
where $\Bar{S}$ denotes the complement of the set $S$. In other words, these are pure multipartite quantum states whose every reduction to a $k-$party state or less is maximally mixed. For such states, the mean-field quantities $m_{\alpha}$ corresponding to the single-particle collective operators, $\hat{S}_{\alpha}$, are such that
\begin{equation}
    m_{\alpha} = \frac{\langle \hat{S}_{\alpha} \rangle}{S} = \sum_{i=1}^{N} \frac{\text{Tr}\left[ \sigma_{\alpha}^{i} {\rho_i}\right]}{S} = 0,
\end{equation}
since the reduced density matrix corresponding to the $i$-th spin, $\rho_i = \text{Tr}_{\Bar{i}}(\rho_{k-\text{uniform}}) = \mathbbm{I}_2$ and $\sigma_{\alpha}^{i}$ are the traceless Pauli operators. Therefore, no dynamical features of such a system can be captured within a mean-field description. 

As an example, we consider the cat state which is a well-known $1$-uniform state, $(\ket{S,S} + \ket{S,-S})/\sqrt{2}$. As $S \to \infty$, under a mean-field approximation, this state completely misses the phase transition into a time crystalline phase as the mean-field quantities remain zero throughout. 
Consequently, the time average of $\langle \hat{S}_z\rangle/S$, an order parameter for the underlying phase transition, also remains zero and featureless. However, the exact solution given by the Lindblad equation records a phase transition independent of the initial state, as long as the state belongs to the subspace with total angular momentum $S$. 
Hence, a mean-field approximation makes it impossible to capture this transition.

This phenomenon can in fact be understood within the Liouville superoperator formalism, where every observable $\mathcal{O}$ at time $t$ can be expressed as
\begin{equation}\label{eq:Timevolobs}
\braket{\hat{\mathcal{O}}(t)}=\sum_{k}e^{(\alpha_k + \iota\beta_k)t}\lbra{l_k}\llket{\rho_0}\lbra{\hat{\mathcal{O}}}\llket{r_k}.
\end{equation} 
Accordingly, the validity of the mean-field approximation relies on the above sum registering non-zero values at time $t=0$ in the thermodynamic limit for a given observable and initial state. 
This could be a function of vanishing overlaps of the observable with individual right eigenvectors or simply a matter of the total sum vanishing. 
As we have seen above, this sum vanishes for $k-$uniform states resulting in the failure of mean-field approximations for these initial states. 
In order to reasonably capture the signatures of the time crystalline dynamics in these systems, one needs to go beyond mean-field theory and incorporate the contribution of the non-zero leading order $n-$particle observables into our description of the system. We do this systematically using cumulant expansion theory that goes beyond the mean-field approximation, yet casts the problem of tracking the dynamics into solving a solvable set of coupled differential equations.

\section{Cumulant Expansion Theory}\label{sec: cumulants}
Cumulant expansions can be thought of as improvements to \sout{the} mean field theory where higher order correlations captured by the few-body observables can be systematically incorporated. 
Such an approximation helps cast the evolution of the few-body observables as a set of coupled equations that follow the BBGKY hierarchy, where the dynamics of the mean of $n$-particle operators depend on the mean of $(n + 1)$-particle operators~\cite{Kubo_original, kira2011semiconductor}. 

The higher order correlations can in general be neglected, enabling a truncation of the BBGKY hierarchy. 
This is especially true for few-body Hamiltonians, where higher-order correlations beyond a certain order are irrelevant to the description of the system, thereby justifying such a truncation~\cite{PhysRevA.82.033810, sanchez2020cumulant, hoffmann2019benchmarking, SHUNJIN1985328}. 
However, to develop a theory that accurately captures the dynamics of the underlying system, it is important to truncate at an order that respects the conserved quantities intrinsic to the system~\cite{kronke2018born, PhysRevB.85.235121}.  
The cumulant expansion method further ensures that truncation at any convenient order results in a closed set of equations. 
For cumulants up to the $s^{th}$ order, a set of dynamical equations can be written in the form, $\dot{\mathcal{C}}_{\vec{s}} = \Vec{F^{s}}\left(\mathcal{C}_{1}, ... \mathcal{C}_{s}, \mathcal{C}_{s+1}\right)$, where $\vec{s} = \{1,2, ... s\}$ is a set that labels the cumulants. 
$\mathcal{C}_{s}$ above stands for an $s^{th}$ order cumulant and $\Vec{F^{s}}$ symbolizes the set of equations for the first $s$ cumulants.  

\begin{figure*}[htp!]
\centering
    \includegraphics[width=\textwidth]{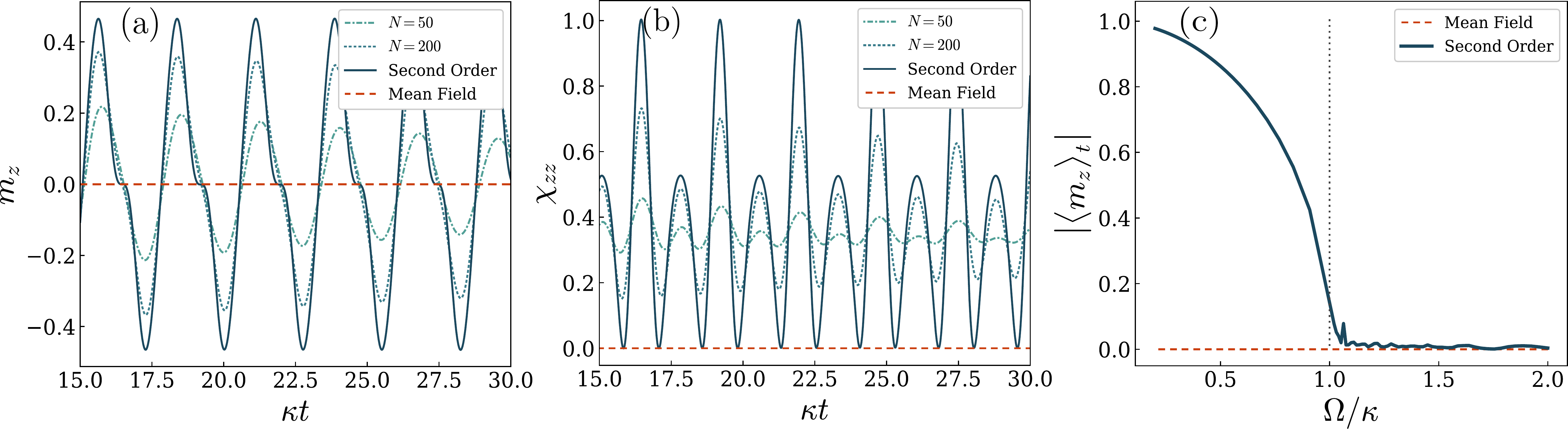}
    \caption{Time series of (a) $m_z$ and (b) $\chi_{zz}$ computed from the exact dynamics (for $N = \{50, 200\}$), mean-field approximation (dashed red line) and the second order cumulant equations (solid dark line) when $\Omega/\kappa = 2.5$ and initialized in the cat state, $\ket{\text{cat}^{(+)}}$. (c) Long-time average of $m_z$ time series as a function of $\Omega/\kappa$ at the level of mean-field approximation and the second order cumulant. The dotted vertical line at $\Omega/\kappa = 1$ represents critical point of the phase transition. The mean-field predictions remain zero throughout in all three figures, indicating its inability to capture the exact dynamics while the second order cumulant captures this transition.}
    \label{fig:figure1}
\end{figure*}

Of particular interest to this study are the first three cumulants, denoted as $\mathcal{C}_{1}^{i} = m_{i}$, $\mathcal{C}_{2}^{ij} = \chi_{ij}$, and $\mathcal{C}_{3}^{ijk} = \tau_{ijk}$, where cumulants are defined in terms of the collective operators $\hat{S}_i$,  for $i,j,k \in \{x, y, z\}$. 
We note that our definition of second-order cumulants correspond to that of the standard definition of equal-time spatial correlation functions in spin systems (see Appendix~\ref{app: cumulants equivalence}). 
Since collective spins scale extensively in the system size, we normalize them using the total spin $S$, such that the first and second order cumulants take the form $m_i=\braket{{\hat{S}_i}}/{S}$ and $\chi_{ij} = \big(\braket{\hat{S}_i\hat{S}_j} - \braket{\hat{S}_i}\braket{\hat{S}_j}\big)/{S^2}$, respectively. 
The third order cumulant has the form $\tau_{ijk} = \big(\braket{\hat{S}_i\hat{S}_j\hat{S}_k} - \braket{\hat{S}_i\hat{S}_j}\braket{\hat{S}_k} - \braket{\hat{S}_j\hat{S}_k}\braket{\hat{S}_i} - \braket{\hat{S}_i\hat{S}_k}\braket{\hat{S}_j} + 2\braket{\hat{S}_i}\braket{\hat{S}_j}\braket{\hat{S}_k}\big)/S^3$, which we set to zero for the purposes of our study. 
On solving for the expectation values of the relevant observables using the Heisenberg equation of motion and substituting $\braket{\hat{S}_i\hat{S}_j\hat{S}_k} \approx \braket{\hat{S}_i\hat{S}_j}\braket{\hat{S}_k} + \braket{\hat{S}_j\hat{S}_k}\braket{\hat{S}_i} + \braket{\hat{S}_i\hat{S}_k}\braket{\hat{S}_j} - 2\braket{\hat{S}_i}\braket{\hat{S}_j}\braket{\hat{S}_k}$, we get a closed set of equations given by,
\begin{subequations}\label{eq:cumulants}
\begin{align}
    \dot{m}_x &= \kappa\chi_{xz} + \kappa m_{x} m_{z}\label{eq:ord2_start}\\
    \dot{m}_{y} &=  \kappa\chi_{yz} + (\kappa m_{y} -  \omega_{0})m_{z}\\
    \dot{m}_{z} &= -  \kappa\chi_{xx} -  \kappa\chi_{yy} - \kappa m_{x}^{2} - \kappa m_{y}^{2} +  \omega_{0}m_{y}\\
    \dot{\chi}_{xx} &= 2  \kappa m_{z}\chi_{xx} + 2  \kappa m_{x}\chi_{xz}\\
    \dot{\chi}_{xy} &= 2  \kappa m_{z}\chi_{xy} +  (\kappa m_{y} - \omega_{0})\chi_{xz} + \kappa m_{x}\chi_{yz} \\
    \dot{\chi}_{xz} &= - 2\kappa m_{x}\chi_{xx}  + (\omega_{0} - 2 \kappa m_{y})\chi_{xy} \nonumber\\
    &\hspace{36mm} +   \kappa m_{z}\chi_{xz} +  \kappa m_{x}\chi_{zz} \\
    \dot{\chi}_{yy} &= 2  \kappa m_{z}\chi_{yy} + (2 \kappa m_{y} - 2 \omega_{0})\chi_{yz} \\
    \dot{\chi}_{yz} &= - 2  \kappa m_{x}\chi_{xy} + (\omega_{0} - 2 \kappa m_{y})\chi_{yy}  \nonumber\\
    &\hspace{25mm} +   \kappa m_{z}\chi_{yz} +   (\kappa m_{y} - \omega_{0})\chi_{zz}\\
    \dot{\chi}_{zz} &= - 4  \kappa m_{x}\chi_{xz} + (2 \omega_{0} - 4  \kappa m_{y})\chi_{yz}.
    \label{eq:ord2_end}
\end{align}
\end{subequations}

The second order cumulants above can be set to zero to recover the widely studied mean field equations given in Appendix~\ref{app: meanfieldeqn} for pedagogical completeness. 
In general, such an approximation works pretty well in the thermodynamic limit for interacting models above the critical dimension, where the few-body interactions die down within a finite distance. 
However, the model under consideration is zero-dimensional and as such there is little reason to specify a cutoff to truncate the cumulants, forcing us to deal with a set of countably infinite equations.
Therefore, we take a more pragmatic approach and ask the question as to what the lowest order of truncation can possibly be. We conclude that since the strong symmetry $S^2$ can be written in the form,
\begin{equation}
    S^2 = m_x^2 + m_y^2 + m_z^2 + \chi_{xx} + \chi_{yy} + \chi_{zz},
\end{equation}
second order cumulants constitute the lowest order of truncation that respects the strong symmetry of the system. 
As is easy to see, the presence of second order cumulants in a strong symmetry of the system leaves the mean field insufficient to describe the evolution of those states with non-zero initial cumulants. 
In the following section, we show that the second-order cumulants show qualitatively different behavior and capture several important dynamical features missed by the mean field.

\section{Results} \label{sec: results}

\begin{figure*}
    \includegraphics[width=\textwidth]{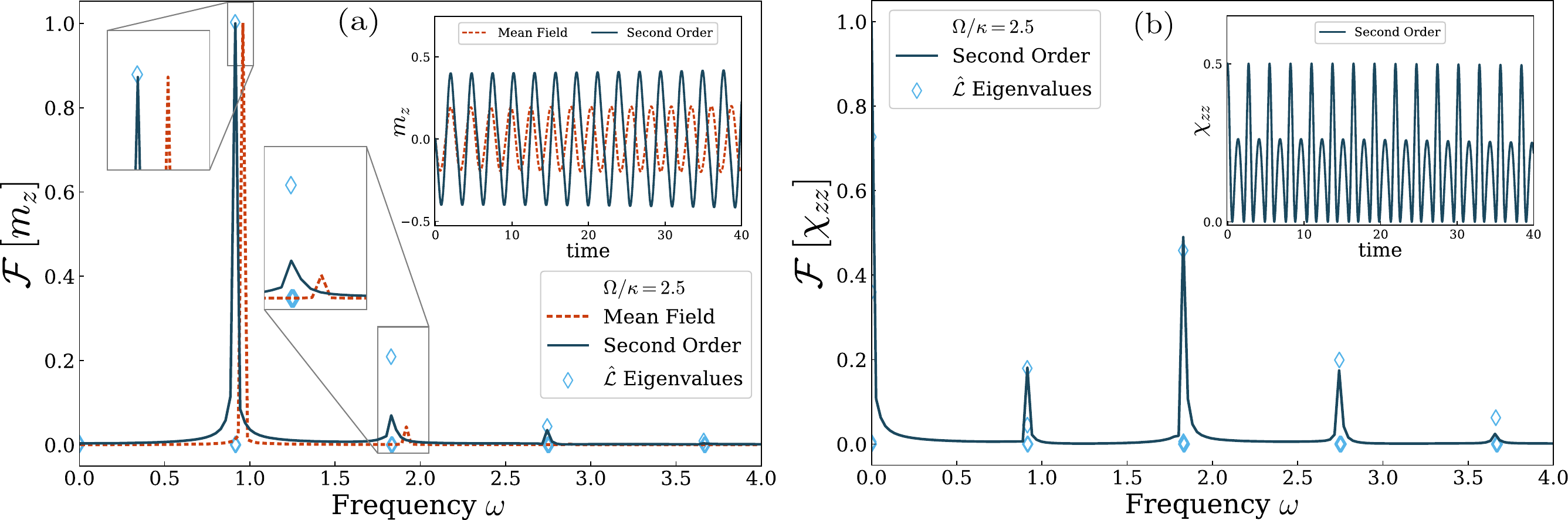}
    \caption{(a) Normalized Fourier transform of the time series of $m_z$ given by mean-field (dashed red line) and second order cumulant (solid dark line) equations for the initial state $\ket{\Psi\left(\pi/4,0 \right)}$. The blue diamonds are plotted at frequencies given by the asymptotic scaling of the imaginary parts of the eigenvalues of the Liouvillian and the heights at which they are plotted are given by the products of the overlaps of the right and left eigenvectors with $S_z/S$ and the initial state respectively for $N = 60$ as suggested in Eq.~\ref{eq:Timevolobs}. The inset shows the time evolution of $m_z$ as per the mean-field (dashed red line) and second order cumulant (solid dark line) equations. (b) Normalized Fourier transform of $\chi_{zz}$ time series obtained from the second order cumulant equations for the initial state $\ket{\Psi\left(\pi/4,0 \right)}$. The blue diamonds are the eigenfrequencies obtained using the finite size scaling and the heights at which they have plotted have a similar meaning as in (a). The inset shows the corresponding time series. The system is in a time crystalline phase with $\omega/\kappa = 2.5$ in both the cases.}
    \label{fig:fig_2}
\end{figure*}

As stated above, the mean field successfully captures the dynamics for coherent states and product initial states such as $|S, S\rangle$, whereas for correlated initial states this approximation fails as it involves neglecting higher-order correlations. There is an initial state dependence in the dynamics of these observables for generic initial states, as seen from Eq.~(\ref{eq:Timevolobs}), which becomes important as we consider states with non-vanishing higher order cumulants.
In such cases, the cumulant hierarchy successfully emulates the exact dynamics while the mean field equations fail to capture crucial dynamical features. Here, we demonstrate two specific examples where the second order cumulant equations (Eq.~5) better capture the dynamics than the mean-field.

First, we re-examine the even parity cat state, $\ket{\text{cat}^{(+)}}=(\ket{S,S}+\ket{S,-S})/\sqrt{2}$. 
As discussed in Section~\ref{sec: limits}, this is a $1$-uniform state and hence the mean-field variables $(m_x, m_y, m_z)$ are all initially zero. 
However, on the inclusion of the second order cumulant, we notice that there is one non-zero variable in the thermodynamic limit, that is, $\chi_{zz}=1$ such that the strong symmetry $S^2$ is preserved. 
While the mean field completely misses the time crystalline behavior, we note that on resorting to second order cumulant equations, it is captured to a high degree of accuracy. 
This is shown in Fig.~\ref{fig:figure1} through time-series plots for $m_z$ and $\chi_{zz}$. Consequently, the phase transition that is missed on employing a mean-field approximation is also to a good approximation captured on including the second order cumulants. 
This is shown in Fig.~\ref{fig:figure1}(c) where the time average of $\langle \hat{S}_z\rangle/S$ is used as the order parameter for the second-order phase transition.

In contrast to the $1$-uniform state where all the mean-field variables vanish, we now examine a case where not all mean-field variables are initially zero. To this end, we study a class of initial states, namely arbitrary superposition of spin coherent states. 
These entangled states can be experimentally prepared in various quantum systems including atom-cavity systems, circuit QED systems, Bose-Einstein condensates and thermal Rydberg atoms~\cite{PhysRevA.57.2247, PhysRevLett.113.090401, signoles2014confined}.  
They have also been shown to be great candidates for high-precision phase estimation protocols that beat the standard quantum limit while remaining robust against detection noise and dissipation~\cite{huang2015quantum, PhysRevA.98.012129,PhysRevA.86.033823, PhysRevA.104.053712}.  
The higher order cumulants are non-zero for this class of states and hence we study the dynamics starting from these correlated initial states using Eq.~(\ref{eq:cumulants}). Through this, we demonstrate the inaccuracy of mean-field predictions by comparing it against the exact dynamics. 
The spin coherent state is defined as a direct product of the individual qubit states,
\begin{equation}
    \ket{\theta, \phi} = \bigotimes_{i} \left[\cos\left( \frac{\theta}{2}\right) \ket{0}_i + e^{\iota \phi}\sin\left( \frac{\theta}{2}\right) \ket{1}_i\right].
\end{equation}  
This state can be thought of as a superposition of Dicke states as follows,
\begin{align}
    \ket{\theta,\phi} = \sum_{m = -J}^{J} c_m(\theta) e^{-\iota(J+m)\phi}\ket{J,m},\label{eq:spin_coherent_state}
\end{align}
where, $c_m(\theta) = [(2J)!/\{(J+m)!(J-m)!\}]^{1/2}\sin^{J+m}\left({\theta}/{2}\right)\cos^{J-m}\left({\theta}/{2}\right)$. 
The superposition of these spin coherent states can now be defined as $\ket{\Psi(\theta,\phi)} =(\ket{\theta,\phi}+\ket{\pi-\theta,\phi})/{\sqrt{2}}$. The cat state defined above is a special case of this state when $\theta = \phi = 0$.
However, unlike the cat state, $\ket{\Psi(\theta,\phi)}$ is in general a superposition of two states that are not mutually orthogonal to each other. 

We consider this aforementioned state for $\theta = \pi/4$ and $\phi = 0$ for which the initial cumulants up to second order are all zero in the thermodynamic limit, except for $m_x=1/\sqrt{2}$ and $\chi_{zz}=1/2$. 
Unlike the case of the cat state before, a first order cumulant is non-zero here. Through this example, we show that mean-field can produce dynamics with quantitatively different behavior from the exact solution. 
We analyze the time series and perform a Fourier transform of the dynamics generated by the mean-field equations (Eq.~\ref{eq: meanfield}) and the cumulant equations (Eq.~\ref{eq:cumulants}) and compare them against the expected frequencies as given by the Liouville superoperator theory. 
Since the mean-field and cumulant equations are derived for the limit $N \to \infty$, we perform a finite size scaling of the Liouville superoperator eigenvalues in order to nullify any finite size contribution to the eigenvalues. 
For this, we follow the procedure in~\cite{hajdusek_seeding_2022} and assume a fitting function of the form $\lambda(N) = \sum_{i = 0}^{\mu} a_{i}/N^i$ for the real and imaginary parts of the eigenvalues such that $\lambda(N) \to a_0$ in the thermodynamic limit. 
This scaling allows us to choose those eigenvalues for which the imaginary parts are non-zero and the corresponding real parts are zero such that they can potentially show up in the dynamics of observables in the thermodynamic limit.

Fig.~\ref{fig:fig_2}(a) shows the Fourier analysis for $m_z$ evolution both at the level of mean-field approximation and the second order cumulant expansion, along with the eigenfrequencies obtained on fitting the eigenvalues. 
We see that the mean-field predictions of the Fourier components deviate from the exact values as given by the Liouville eigenspectrum whereas the Fourier transform for $m_z$ dynamics as given by the second-order cumulant almost exactly captures it. 
This can also be seen from the inset to Fig.~\ref{fig:fig_2}(a) where the time evolution given by the mean-field equations differ significantly from that given by the second order cumulant equations. 
Fig.~\ref{fig:fig_2}(b) further suggests a high degree of accuracy between the Fourier transform of the $\chi_{zz}$ time evolution given by the second order cumulant equation and the eigenfrequencies as predicted by the asymptotic Liouvillian fit discussed above.

\section{Discussion \& Conclusion} \label{sec: conclusion}

Continuous time crystals are phases of matter that have improved our understanding of exotic phase transitions. They have recently been proposed as sensors and heat engines~\cite{carollo_nonequilibrium_2020, montenegro2023quantum, paulino_nonequilibrium_2023, PhysRevLett.132.050801}, making thorough understanding of their dynamics be of both theoretical as well as practical value.
We investigated continuous time crystals initialised with correlated states. We showed that the mean-field approximation failed to capture the dynamics of these systems for such initial states. Further, we used a cumulant expansion method to systematically incorporate the effects of correlations up to second order, which then efficiently captured the exact dynamics.

A natural question that might arise is what justifies the truncation at the second order. 
We note that the second order cumulant is the smallest order that captures the symmetries of the Liouville superoperator. 
Furthermore, the presented examples show that the second order constitutes the lowest non-zero leading order cumulant irrespective of how correlated the initial states are. 
This can also be seen from the fact that the squares of the collective operators contain traceful on-site interaction terms, leading to a finite contribution in the cumulant hierarchy that can then capture the essential features of the system dynamics. 
We explain the initial state dependence of the mean-field by noting that the sum of the overlaps between the observable and the right eigenvectors as given in Eq.~(\ref{eq:Timevolobs}) can be zero for all the single particle observables, causing them to not evolve. 
In other examples, we noted that the deviation of the mean-field solution from that of the exact solution is due to the presence of higher order correlations.

While mean-field is thought to be exact in the thermodynamic limit, many-body physics presents us with several cases where it fails. 
In the case of many-to-one models like the central spin models, the validity of mean-field theory depends on the scaling of model parameters~\cite{PhysRevResearch.5.033148}. Here, truncating the cumulant expansion at appropriate orders was shown to help regain the steady state predicted by the exact dynamics. 
There are several instances where cumulant expansions become important to recover the exact dynamical features. This  includes light-matter interaction models, especially in the low photon limit where fluctuations play a major role in the system dynamics~\cite{PhysRevA.82.033810}. 
The current work adds to this analysis by highlighting the insufficiency of mean-field theory to study specific examples of continuous time crystalline behavior. In addition, alongside approximation techniques like the spin wave theory, we hope that cumulant expansion theories have a role to play in examining the robustness of continuous time crystals. Finally, we note that our method is adjacent to the method of fluctuation operators studied in the context of CTCs \cite{goderis1989non, goderis1990dynamics, PhysRevA.105.L040202, PhysRevA.106.012212, mattes_entangled_2023}.

\section{Acknowledgements}\label{sec: acknowledgements}
SV acknowledges support
from Government of India DST-QUEST grant number DST/ICPS/QuST/Theme-4/2019. YI acknowledges Government of India DST-INSPIRE for a fellowship. MH was supported
by [MEXT Quantum Leap Flagship Program] Grants No.
JPMXS0118067285 and No. JPMXS0120319794.
The authors thank Giulia Piccitto, Himadri Shekhar Dhar and Arul Lakshminarayan for discussions.

\clearpage
\appendix
\maketitle
\onecolumngrid
\section{Mean-Field Equations} \label{app: meanfieldeqn}
The mean field equations for the canonical model of boundary time crystals can be regained by setting the second order cumulants to zero in Eq~\ref{eq:cumulants}. We explicitly write down the equations for the mean-field variables below as,
\begin{subequations} \label{eq: meanfield}
\begin{align}
    \dot{m}_x &= \kappa m_{x} m_{z}\label{eq:ord1_start}\\
    \dot{m}_{y} &=  (\kappa m_{y} -  \omega_{0})m_{z}\\
    \dot{m}_{z} &= - \kappa m_{x}^{2} - \kappa m_{y}^{2} +  \omega_{0}m_{y}
\end{align}
\end{subequations}

\section{Modified Dynamical Equations Upon Truncation at the Third-order Cumulants}
The equations for the second order cumulant evolution presented in the main text pertained to the approximation where all moments higher than two are zero. Here, the modified second order cumulant equations in the presence of higher-order cumulants is presented.
\begin{subequations}
\begin{align}
\dot{\chi}_{xx}&=
2 \chi_{xx} \kappa m_{z} + 2 \chi_{xz} \kappa m_{x} + 2 \kappa \tau_{xxz}\\
\dot{\chi}_{yy}&=
2 \chi_{yy} \kappa m_{z} + 2 \chi_{yz} \kappa m_{y} - 2 \chi_{yz} \omega_{0} + 2 \kappa \tau_{yyz}\\
\dot{\chi}_{zz}&=
- 4 \chi_{xz} \kappa m_{x} - 4 \chi_{yz} \kappa m_{y} + 2 \chi_{yz} \omega_{0} - 2 \kappa \tau_{xxz} - 2 \kappa \tau_{yyz}\\
\dot{\chi}_{xy}&= \dot{\chi}_{yx} = 
2 \chi_{xy} \kappa m_{z} +  \chi_{xz} \kappa m_{y} - \chi_{xz} \omega_{0} +  \chi_{yz} \kappa m_{x} + 2 \kappa \tau_{xyz}\\
\dot{\chi}_{xz}&= \dot{\chi}_{zx} = 
- 2 \chi_{xx} \kappa m_{x} - 2 \chi_{xy} \kappa m_{y} + \chi_{xy} \omega_{0} +  \chi_{xz} \kappa m_{z} +  \chi_{zz} \kappa m_{x} -  \kappa \tau_{xxx} -  \kappa \tau_{xyy} +  \kappa \tau_{xzz}\\
\dot{\chi}_{yz}&= \dot{\chi}_{zy} = 
- 2 \chi_{xy} \kappa m_{x} - 2 \chi_{yy} \kappa m_{y} + \chi_{yy} \omega_{0} +  \chi_{yz} \kappa m_{z} +  \chi_{zz} \kappa m_{y} - \chi_{zz} \omega_{0} -  \kappa \tau_{xxy} -  \kappa \tau_{yyy} +  \kappa \tau_{yzz}
\end{align}
\end{subequations}
The equations for the third order cumulants are as follows.
\begin{equation}
\begin{split}
\dot{\tau}_{xxx}&=
6 \chi_{xx} \chi_{xz} \kappa + 3 \kappa m_{x} \tau_{xxz} + 3 \kappa m_{z} \tau_{xxx}\\
\dot{\tau}_{yyy}&=
6 \chi_{yy} \chi_{yz} \kappa + 3 \kappa m_{y} \tau_{yyz} + 3 \kappa m_{z} \tau_{yyy} - 3 \omega_{0} \tau_{yyz}\\
\dot{\tau}_{zzz}&=
- 6 \chi_{xz}^{2} \kappa - 6 \chi_{yz}^{2} \kappa - 6 \kappa m_{x} \tau_{xzz} - 6 \kappa m_{y} \tau_{yzz} + 3 \omega_{0} \tau_{yzz}\\
\dot{\tau}_{xxy} &= \dot{\tau}_{xyx} = \dot{\tau}_{yxx} =
2 \chi_{xx} \chi_{yz} \kappa + 4 \chi_{xy} \chi_{xz} \kappa + 2 \kappa m_{x} \tau_{xyz} +  \kappa m_{y} \tau_{xxz} + 3 \kappa m_{z} \tau_{xxy} - \omega_{0} \tau_{xxz}\\
\dot{\tau}_{xxz}& = \dot{\tau}_{xzx} = \dot{\tau}_{zxx} =
- 2 \chi_{xx}^{2} \kappa + 2 \chi_{xx} \chi_{zz} \kappa - 2 \chi_{xy}^{2} \kappa + 2 \chi_{xz}^{2} \kappa - 2 \kappa m_{x} \tau_{xxx} + 2 \kappa m_{x} \tau_{xzz} - 2 \kappa m_{y} \tau_{xxy} + 2 \kappa m_{z} \tau_{xxz} + \omega_{0} \tau_{xxy}\\
\dot{\tau}_{xyy}&= \dot{\tau}_{yxy} = \dot{\tau}_{yyx} = 
4 \chi_{xy} \chi_{yz} \kappa + 2 \chi_{xz} \chi_{yy} \kappa +  \kappa m_{x} \tau_{yyz} + 2 \kappa m_{y} \tau_{xyz} + 3 \kappa m_{z} \tau_{xyy} - 2 \omega_{0} \tau_{xyz}\\
\dot{\tau}_{zyy}&= \dot{\tau}_{yzy} = \dot{\tau}_{yyz} = 
- 2 \chi_{xy}^{2} \kappa - 2 \chi_{yy}^{2} \kappa + 2 \chi_{yy} \chi_{zz} \kappa + 2 \chi_{yz}^{2} \kappa - 2 \kappa m_{x} \tau_{xyy} - 2 \kappa m_{y} \tau_{yyy}\nonumber\\ 
&\hspace{40mm} + 2 \kappa m_{y} \tau_{yzz} + 2 \kappa m_{z} \tau_{yyz} + \omega_{0} \tau_{yyy} - 2 \omega_{0} \tau_{yzz}\\
\dot{\tau}_{xzz}& = \dot{\tau}_{zxz} = \dot{\tau}_{zzx} = 
- 4 \chi_{xx} \chi_{xz} \kappa - 4 \chi_{xy} \chi_{yz} \kappa + 2 \chi_{xz} \chi_{zz} \kappa - 4 \kappa m_{x} \tau_{xxz} +  \kappa m_{x} \tau_{zzz} - 4 \kappa m_{y} \tau_{xyz} +  \kappa m_{z} \tau_{xzz} + 2 \omega_{0} \tau_{xyz}\\
\dot{\tau}_{yzz}&= \dot{\tau}_{zyz} = \dot{\tau}_{zzy} =  - 4 \chi_{xy} \chi_{xz} \kappa - 6 \chi_{yy} \chi_{yz} \kappa + 2\chi_{yy} \chi_{zy} \kappa + 2 \chi_{yz} \chi_{zz} \kappa - 2 \chi_{yz} \kappa m_{x}^{2}\nonumber\\ 
&\hspace{30mm} - 2 \chi_{yz} \kappa m_{y}^{2} + 2 \chi_{yz} m_{y} \omega_{0} + 2 \chi_{zy} \kappa m_{x}^{2} + 2 \chi_{zy} \kappa m_{y}^{2} - 2 \chi_{zy} m_{y} \omega_{0} - 4 \kappa m_{x} \tau_{xyz} \nonumber\\ 
&\hspace{40mm}- 4 \kappa m_{y} \tau_{yyz} +  \kappa m_{y} \tau_{zzz} +  \kappa m_{z} \tau_{yzz} + 2 \omega_{0} \tau_{yyz} - \omega_{0} \tau_{zzz}\\
\dot{\tau}_{xyz} &= \dot{\tau}_{xzy} = \dot{\tau}_{yxz} = \dot{\tau}_{yzx} = \dot{\tau}_{zxy} = \dot{\tau}_{zyx} = 
- 2 \chi_{xx} \chi_{xy} \kappa - 2 \chi_{xy} \chi_{yy} \kappa + 2 \chi_{xy} \chi_{zz} \kappa + 2 \chi_{xz} \chi_{yz} \kappa   \nonumber\\ 
&\hspace{30mm} - 2 \kappa m_{x} \tau_{xxy} +  \kappa m_{x} \tau_{yzz} - 2 \kappa m_{y} \tau_{xyy} +  \kappa m_{y} \tau_{xzz} + 2 \kappa m_{z} \tau_{xyz} + \omega_{0} \tau_{xyy} - \omega_{0} \tau_{xzz}
\end{split}
\end{equation}

\section{Equivalence of the Definitions of Cumulants}
\label{app: cumulants equivalence} 
We formally prove that in the thermodynamic limit, the definition of second-order cumulants in Section~\ref{sec: cumulants} is equivalent to the standard definition of correlation function. 
\begin{align*}
    \chi_{ij} &= \big(\braket{\hat{S}_i\hat{S}_j} - \braket{\hat{S}_i}\braket{\hat{S}_j}\big)/{S^2}\\
    &= \big(\braket{\sum_{pq}\hat{\sigma}_i^p\hat{\sigma}_j^q} - \braket{\sum_{p}\hat{\sigma}_i^p}\braket{\sum_{q}\hat{\sigma}_j^q}\big)/{S^2}\\
    &= \big(\braket{\sum_{p}\hat{\sigma}_i^p\hat{\sigma}_j^p} + \braket{\sum_{p\neq q}\hat{\sigma}_i^p\hat{\sigma}_j^q} - \braket{\sum_{p}\hat{\sigma}_i^p}\braket{\sum_{q}\hat{\sigma}_j^q}\big)/{S^2}
\end{align*}
Since the system is permutationally invariant, $\braket{\hat{\sigma}_i^p\hat{\sigma}_j^q}$ does not depend on $p,q$. Also, $\hat{\sigma}_i^p\hat{\sigma}_j^p= \iota \epsilon_{ijk}\hat{\sigma}_k^p$.
\begin{align*}
    \chi_{ij} &= \big(\pm\iota\sum_{p}\braket{\hat{\sigma}_k^p} + \sum_{p\neq q}\braket{\hat{\sigma}_i^p\hat{\sigma}_j^q} - \sum_{p}\braket{\hat{\sigma}_i^p}\sum_{q}\braket{\hat{\sigma}_j^q}\big)/{S^2}\\
    &= \big(\pm\iota N\braket{\hat{\sigma}_k^p} + N(N-1)\braket{\hat{\sigma}_i^p\hat{\sigma}_j^q} - N\braket{\hat{\sigma}_i^p}\times N\braket{\hat{\sigma}_j^q}\big)/{S^2}\\
    &= \big(\pm 2\iota S\braket{\hat{\sigma}_k^p} + 2S(2S-1)\braket{\hat{\sigma}_i^p\hat{\sigma}_j^q} - 4S^2\braket{\hat{\sigma}_i^p}\braket{\hat{\sigma}_j^q}\big)/{S^2}\\
    &= 4(\braket{\hat{\sigma}_i^p\hat{\sigma}_j^q} - \braket{\hat{\sigma}_i^p}\braket{\hat{\sigma}_j^q})\ \left (\mathrm{as}\ S\to\infty\right),
\end{align*}
where, $(\braket{\hat{\sigma}_i^p\hat{\sigma}_j^q} - \braket{\hat{\sigma}_i^p}\braket{\hat{\sigma}_j^q})$ is the equal-time correlation function between two variables, $\hat{\sigma}_i$ and $\hat{\sigma}_j$ at arbitrary sites $p$ and $q$.

\section{Cumulants of the Coherent State} \label{app: cum_for_cohstate}
Here, we show that all the cumulants for a spin coherent state are zero in the thermodynamic limit. 
The spin coherent state,
\begin{equation}
    \ket{\theta,\phi} = \sum_{m=-J}^{J} \binom{2J}{J+m}^\frac{1}{2}\sin^{J+m}\left(\frac{\theta}{2}\right)\cos^{J-m}\left(\frac{\theta}{2}\right)e^{-\iota\phi(J+m)}\ket{J,m}\label{eq:spin_coherent_state}.
\end{equation}
In this state, the first order cumulants can be evaluated to obtain,
\begin{align*}
    \frac{\left\langle \hat{S}_x\right\rangle}{S} = \sin\theta\cos\phi \quad 
    \frac{\left\langle \hat{S}_y\right\rangle}{S} = \sin\theta\sin\phi \quad 
    \frac{\left\langle \hat{S}_z\right\rangle}{S} = -\cos\theta.
\end{align*}
The second order cumulant scales as the inverse of the total angular momentum $S$ that they vanish in the thermodynamic limit, as shown below.
\begin{align*}
    \chi_{xx} &= \frac{1}{2S}\left[\cos^4\left(\frac{\theta}{2}\right)+\sin^4\left(\frac{\theta}{2}\right)-\frac{\sin^2\theta\cos2\phi}{2}\right] \\
    \chi_{xy} &= -\frac{1}{2S}\left[\sin^2\theta\sin\phi\cos\phi + \iota \cos\theta\right]\\
    \chi_{xz} &= \frac{\sin\theta}{2S}\left[e^{\iota\phi}\sin^2\left(\frac{\theta}{2}\right)-e^{-\iota\phi}\cos^2\left(\frac{\theta}{2}\right)\right]\\
    \chi_{yy} &= \frac{1}{2S}\left[\cos^4\left(\frac{\theta}{2}\right)+\sin^4\left(\frac{\theta}{2}\right)+\frac{\sin^2\theta\cos2\phi}{2}\right] 
    \\
    \chi_{yz} &= \frac{\sin\theta}{2iS}\left[e^{\iota\phi}\sin^2\left(\frac{\theta}{2}\right)+e^{-\iota\phi}\cos^2\left(\frac{\theta}{2}\right)\right]\\
    \chi_{zz} &= \frac{1}{2S}\sin^2\theta
\end{align*}
Similarly, each higher order of cumulant picks up an extra factor of $1/S$ such that all correlations vanish as $S\to\infty$.

\end{document}